\documentclass[journal=jacsat,manuscript=article]{achemso}

\usepackage{amsmath}
\usepackage{upgreek}

\author{Anastasiia Lukovkina}
\affiliation {Department of Quantum Matter Physics, University of Geneva, CH-1211 Geneva, Switzerland}
\author{Sara A. López-Paz}
\affiliation {Department of Quantum Matter Physics, University of Geneva, CH-1211 Geneva, Switzerland}
\author{Céline Besnard}
\affiliation {Department of Quantum Matter Physics, University of Geneva, CH-1211 Geneva, Switzerland}
\author{Laure Guenee}
\affiliation {Department of Quantum Matter Physics, University of Geneva, CH-1211 Geneva, Switzerland}
\author{Fabian O. von Rohr}
\affiliation {Department of Quantum Matter Physics, University of Geneva, CH-1211 Geneva, Switzerland}
\email{*Correspondance}
\author{Enrico Giannini}
\affiliation {Department of Quantum Matter Physics, University of Geneva, CH-1211 Geneva, Switzerland}

\title {Controlling the Magnetic Properties of the van der Waals Multiferroic Crystals Co$_{1-x}$Ni$_x$I$_2$}

\begin{document}

\begin{tocentry}

\includegraphics[width=8cm]{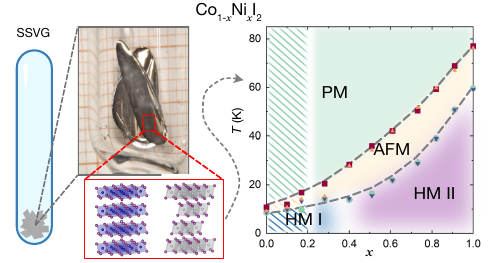}
\centering

\end{tocentry}

\begin{abstract}
  The structurally related compounds NiI$_2$ and CoI$_2$ are multiferroic van der Waals materials, in which helimagnetic orders exist simultaneously with electric polarization. Here, we report on the evolution of the crystal structure and of the magnetic properties across the solid solution Co$_{1-x}$Ni$_x$I$_2$. We have successfully grown crystals of the whole range of the solid solution, i.e. $x = 0 - 1$, by employing the self-selecting vapor growth (SSVG) technique and by carefully tuning the synthesis conditions according to the chemical composition. Our structural investigations show that the crystal symmetry changes from $P\bar{3}m1$ to $R\bar{3}m$ when Ni substitutes for Co beyond $x = 0.2$. Both the lattice parameters and magnetic properties evolve continuously and smoothly from one end member to the other, showing that they can be finely tuned by the chemical composition. We also observe that the Ni substitution degree in the solid solution affects the metamagnetic transition typical for CoI$_2$ at high magnetic fields. In particular, we find the existence of the metamagnetic transition similar to that for CoI$_2$ in the NiI$_2$ structure. Based on magnetic measurements we construct the phase diagram of the Co$_{1-x}$Ni$_x$I$_2$ system. Controlling the magnetic properties by the chemical composition may open new pathways for the fabrication of electronic devices made of two-dimensional (2D) flakes of multiferroic van der Waals materials.
\end{abstract}

\section{Introduction}

The discovery of graphene through the mechanical exfoliation technique has led to a surge in research focused on van der Waals materials.\cite{novoselov2004electric} Van der Waals materials exhibit weak interaction between two-dimensional (2D) layers with strong covalent bonds within them, offering distinctive ways to modify their properties, including layer stacking, rotational adjustments, and the intercalation of different molecular species.\cite{whittingham1978chemistry,hulliger2012structural,manzeli20172d,zheng2021polar} Tailoring these chemical parameters -- from a single layer to multiple layers -- has proven to be a pathway for altering the physical properties. \cite{geim2013van,duong2017van}

Due to the coexistence of van der Waals forces and magnetic interactions, magnetic van der Waals materials have emerged as a class that offers new opportunities, leading to a rich and diverse range of functionalities.\cite{gong2017discovery,huang2017layer,burch2018magnetism,gibertini2019magnetic}
The advent of 2D magnets -- through mechanical exfoliation from bulk van der Waals magnetic materials -- has opened up a new arena for exploring and controlling 2D magnetism and magnetic devices\cite{gibertini2019magnetic,sierra2021van}. This progress is propelling the development of spintronics and magneto-optics applications.\cite{song2018giant,nvemec2018antiferromagnetic} The diverse array of these 2D van der Waals magnets now includes, among others, transition metal halides\cite{wang2019determining}, chalcogenides\cite{bonilla2018strong}, and mixed-anion compounds of transition metals\cite{wu2023gate,telford2020layered}, exemplified by the diverse properties of CrSBr.\cite{wilson2021interlayer,wu2022quasi,lopez2022dynamic}

Type-II multiferroicity has been discovered for some of these compounds.\cite{McGuire2017} This makes these materials promising candidates for the creation of the next-generation of electronics with ferroelectricity existing simultaneously with long-range magnetically ordered structures. If a material has a strong coupling of magnetic and ferroelectric states, it provides an opportunity to tune ferroelectric properties with a magnetic field and vice versa.\cite{Khomskii2009, Tokura2010} 

In the growing family of novel magnetic van der Waals materials, transition metal diiodides, with the transition metals in a +2 oxidation state, have been found as hosts of magnetic excitations and as platform for versatile multiferroics; thus they have attracted significant attention lately.\cite{Son2022,Lebedev2023,Bai2021,Kim2023} 

\begin{figure*} 
    \centering
    \includegraphics[width=16cm]{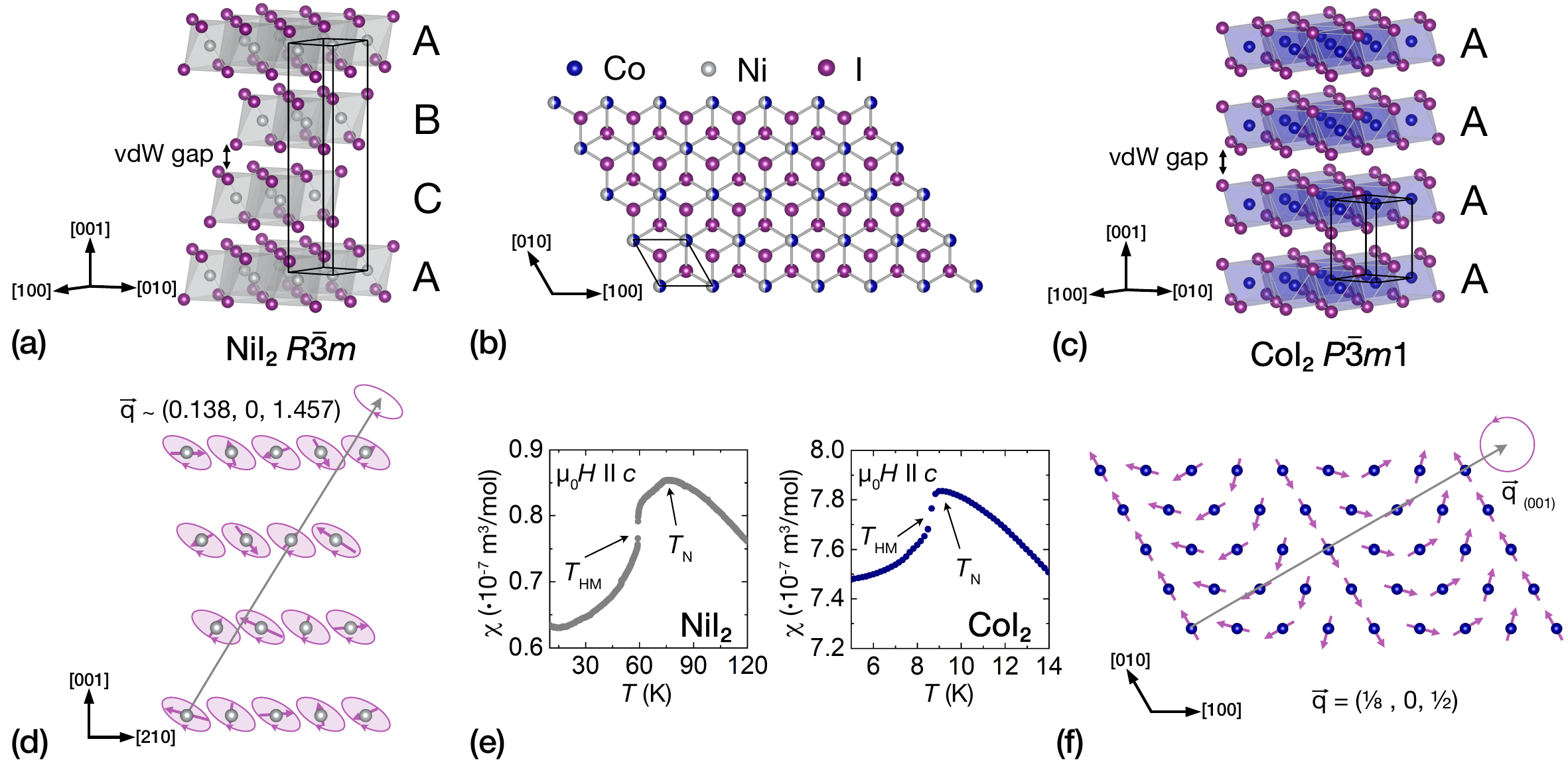}
    \caption{Crystal and magnetic structures of layered van der Waals materials CoI$_2$ and NiI$_2$: (a) Trigonal crystal structure of NiI$_2$ with [ABCA...] stacking. (b) Monolayer of CoI$_2$ or NiI$_2$. The crystal structure of the two materials differs from the stacking of the layers -(I-TM-I)-. (c) Trigonal crystal structure of CoI$_2$ with [AA...] stacking.  (d) Spin arrangement in the helimagnetic ground state of NiI$_2$ with proper screw helical structure. Spins (shown with pink arrows) are lying within the planes perpendicular to the magnetic propagation vector $\Vec{q}$ (coordinates are given in the reciprocal space, direction is shown with a gray arrow), and a translation of the rotation (visualized with pink circles with arrows) occurs in the direction of this vector. (e) Magnetic susceptibility data of the parent compounds NiI$_2$ and CoI$_2$. $T_{\rm N}$ corresponds to the transition to the antiferromagnetic state and $T_{\rm HM}$ to the helimagnetic order. (f) Spin arrangement in the helimagnetic ground state of CoI$_2$ with the cycloid magnetic order. Coordinates of the magnetic propagation vector are given in the reciprocal space, $\Vec{q}_{(001)}$ is a projection of this vector on $(001)$ plane (shown with a gray arrow). The spins (shown with pink arrows) are arranged within the $(001)$ planes and a translation of the rotation (visualized with a pink circuit with the arrow) occurs in the same plane in the direction $\Vec{q}_{(001)}$.}
    \label{fig:CoI2_NiI2}
\end{figure*}

Among them, CoI$_2$ and NiI$_2$ exhibit subtle differences in their crystal structure and slightly different magnetic orders and magnetoelectric coupling.\cite{Kurumaji2013} The same 2D layer of TMI$_2$ (TM = Co, Ni) is stacked  in a [AAA..] sequence in CoI$_2$, and in a in-plane shifted  [ABC...] order in NiI$_2$ as shown in Figure \ref{fig:CoI2_NiI2}(a--c), resulting in space groups $P\Bar{3}m1$ and $R\Bar{3}m$, respectively.\cite{Kuindersma1981}

NiI$_2$ is a type II multiferroic that is a triangular antiferromagnet with a helimagnetic ground state, which exists simultaneously with ferroelectricity.\cite{Kurumaji2013, Song2022} Specifically, NiI$_2$ undergoes two magnetic transitions at low temperatures (Figure \ref{fig:CoI2_NiI2}(e)). The first one is at $T_{\rm N} \approx 76$ K to an antiferromagnetic state with the easy-plane perpendicular to the $c$-axis. The second transition occurs at $T_{\rm HM} \approx 59$ K to a helimagnetic state with an incommensurate proper screw spin order with the magnetic modulation vector $\Vec{q}$ = (0.138,0,1.457).\cite{Billerey1980,Billerey1977, Kuindersma1981,Kurumaji2013} This means that the spins are laying in the planes perpendicular to this vector and the translation of rotation of spins occurs in the direction of the magnetic propagation vector, as visualized in Figure \ref{fig:CoI2_NiI2}(d). Almost at the same temperature, a small distortion occurs in the lattice and the crystal structure becomes monoclinic.\cite{Kuindersma1981} As a consequence of the helimagnetic order, NiI$_2$ exhibits spontaneous polarization, which results in intrinsic ferroelectricity.\cite{Kurumaji2013} Recent reports show that the transition temperature to a ferroelectric order is going down with decreasing thickness of the material but remains finite (21 K) in the monolayer limit.\cite{Ju2021,Song2022, Fumega2022} This makes NiI$_2$ a unique van der Waals compound that displays multiferroic, specifically magnetoelectric, properties down to the monolayer limit.

The van der Waals material CoI$_2$ also hosts a polar order simultaneously with a helimagnetic ground state.\cite{Kurumaji2013} The material undergoes similar transitions: to a long-range antiferromagnetic state at $T_{\rm N} \approx 11$ K and to a helimagnetic state at $T_{\rm HM} \approx 9$ K which coexists with the ferroelectricity (Figure \ref{fig:CoI2_NiI2}(e)).\cite{Kurumaji2013} In CoI$_2$, the helimagnetic cycloid ground state is commensurate with the atomic unit cell with a magnetic propagation vector $\Vec{q} = (\frac{1}{8}, 0, \frac{1}{2})$.\cite{Kuindersma1981, Kim2023} Furthermore, it has been reported that this cycloid coexists with another cycloid magnetic structure with a propagation vector $\Vec{q} = (\frac{1}{12}, \frac{1}{12}, \frac{1}{2})$.\cite{Mekata1992, Kurumaji2013} If one considers only the first cycloid, which is consistent with all reports, it exhibits a spin order where the spins are arranged within the easy $(001)$-plane and the translation of the rotation occurs in the same plane as visualized in Figure \ref{fig:CoI2_NiI2}(f). Moreover, it is worth noticing that CoI$_2$ is a rare example of a triangular Kitaev magnet, where bond-dependent anisotropy of exchange interactions is the origin of the frustrated helimagnetic state. \cite{Kim2023} 

Here, we report on the crystal growth of the whole Co$_{1-x}$Ni$_x$I$_2$ solid solution and on the chemically precise control over structural and the magnetic properties in this van der Waals helimagnetic system. Processing sizeable and pure crystals of all compositions across the solid solution was challenging. We find that physical vapour transport (PVT) growth leads to phase separation in the solid solution, while large homogeneous single crystals could be obtained by using self-selecting vapor growth (SSVG).\cite{Szczerbakow2005} SSVG is a synthesis method that has previously been applied for crystal growth of transition metals chalcogenides, as well as chromium and ruthenium trichlorides and tribromides.\cite{Szczerbakow2005,Yan2023} Here we expand this technique to transition metal diiodides. The substitution of Co for Ni is accompanied by progressive change of both the lattice parameters, as well as the magnetic properties. This emphasizes the potential for tailoring of the long-range magnetic properties via compositional chemistry. The realization of single crystals of the whole solid solution of Co$_{1-x}$Ni$_x$I$_2$ allows us to determine the structural properties and the magnetic phase diagram of this system. Given the multiferroic nature of the parent compounds, it is a plausible assumption that the solid solution may also exhibit multiferroic properties. This is supported by the continuous change of the structural and magnetic properties and the observation of the two characteristic magnetic transitions ($T_{\rm N}$ and $T_{\rm HM}$) for all compositions of the solid solution. Our results indicate that the Co$_{1-x}$Ni$_x$I$_2$ system holds the promise to be a new pathway for the chemically precise fabrication of multiferroic devices made of 2D flakes of this van der Waals material.

\section{Experimental Methods}

\subsection{Synthesis}
For crystal growth, CoI$_2$ (powder, Alfa Aesar, 99,6\%) and NiI$_2$ (powder, Alfa Aesar, 99,6\%) were used as starting materials. They were purified by PVT in temperature gradients of $575\to450$ \textdegree{}C and $700\to500$ \textdegree{}C, respectively. All handling of materials, including before sealing and after opening the quartz ampules, was conducted in a glovebox under argon atmosphere ($<0.1$ ppm of O$_2$, $<0.1$ ppm of H$_2$O). This precaution was essential as the starting materials as well as the grown crystals are highly sensitive to air, and are decomposing rapidly when exposed to moisture. For the synthesis of the Co$_{1-x}$Ni$_x$I$_2$ solid solution, two vapor growth techniques were investigated:

\textbf{Physical Vapor Transport (PVT) Method}\\
For the synthesis of the Co$_{1-x}$Ni$_x$I$_2$ solid solution, we first applied the PVT technique using a horizontal temperature gradient in a tubular furnace. We prepared the starting materials by sealing powders in quartz tubes with an excess of NiI$_2$. We carefully controlled the temperature gradients inside the furnace, which were adjusted according to the specific composition of each sample (see results section for further details). 

\textbf{Self-Selecting Vapor Growth (SSVG) Method} \\
We used the SSVG technique, as implemented in a box furnace with slow cooling, to grow single crystals of the Co$_{1-x}$Ni$_x$I$_2$ solid solution. For the SSVG, powders were sealed in quartz tubes using stoichiometric amounts specific to the targeted stoichiometry. To create the small temperature gradient essential for SSVG we positioned the box furnace on its side, thus establishing a horizontal temperature gradient created by a ventilation hole in the furnace, as suggested by Yan et al.\cite{Yan2023} The quartz tubes were placed vertically in Al$_2$O$_3$ boats, with their height adjusted by firebricks. We experimented with various positions inside a furnace to find the optimal temperature gradient conditions for crystal growth. The tubes were dwelled for 6 hours at the initial temperature before slow cooling. The cooling rates in SSVG were varied from 0.4 \textdegree C  to 1 \textdegree C per hour (see results section for further details). 

\subsection{SEM and Elemental Analysis}
The microstructure of grown crystals was studied with a scanning electron microscope (SEM) JEOL JSM-IT800. The composition of the crystals was determined using energy dispersive X-ray spectroscopy (EDS), with an accelerating voltage of 15 kV. About 5 crystals (pieces of crystals) were tested at least at 10 points in each batch. The average value of Ni substitution in Co$_{1-x}$Ni$_x$I$_2$ and standard deviations were calculated for crystals from one growth. As Co and Ni diiodides are semiconducting, thin crystals were used to avoid charging, and they were cleaved directly before measurements to minimize interactions with moisture from air.

\subsection{Phase Analysis}
X-ray diffraction was collected in Bragg-Brentano reflection geometry on single thin plate-like crystals, exposing the $(001)$ plane perpendicular to the scattering vector, in a PANalytical Aeris diffractometer with Cu-$K_\alpha$ radiation. We have used sample holders for air sensitive samples in which the crystal is protected by a Kapton foil against reactions with air.

Powder X-ray diffraction (PXRD) was measured on a Rigaku SmartLab diffractometer in capillary mode with Cu-$K_\alpha$ radiation. The crystals were finely ground with starch in an agate mortar to decrease the preferred orientation of the strongly layered material. The powders were sealed in glass capillaries with an inner diameter of 0.5 mm. 

Single crystal X-ray diffraction (SXRD) was performed under perfluorinated oil and nitrogen flow with Rigaku XtaLab Synergy single crystal X-ray diffractometer equipped with a HiPix detector and Mo-$K_\alpha$ radiation using a cryostream at $T = 120$ K.  

\subsection{Magnetic Properties}
The temperature-dependent and field-dependent magnetic moment was measured using Physical Property Measurements in a cryogen-free system (PPMS DynaCool from Quantum Design) equipped with a vibrating sample magnetometry (VSM) option and a 9 T magnet. Crystals were tightly sandwiched between two pieces of Kapton tape in an argon glovebox in order to avoid interactions with moisture or air. The magnetic field for measurements was applied in two orientations: parallel and perpendicular to the crystallographic $c$-axis. 

\section{Results and Discussion}

\subsection{Crystal Growth}

Single crystals of the end members of the solid-solution, namely CoI$_2$ and NiI$_2$, have previously been grown using physical vapor transport (PVT) and the Bridgman method.\cite{Kuindersma1981, Kurumaji2013, Son2022} We have reproduced the crystal growth results using PVT. In our experiments, the optimized temperature gradients were $550\to450$ \textdegree C and $640\to500$ \textdegree C for growing crystals of CoI$_2$ and NiI$_2$, respectively. The differing vapor pressures of CoI$_2$ and NiI$_2$ are likely the reason for the different PVT conditions for growing crystals of the two end compounds. Single crystals grown by this method were as large as  $6.0\times6.0\times0.1$ mm$^3$. 

\begin{figure*} [h]
    \centering
    \includegraphics[width=\textwidth]{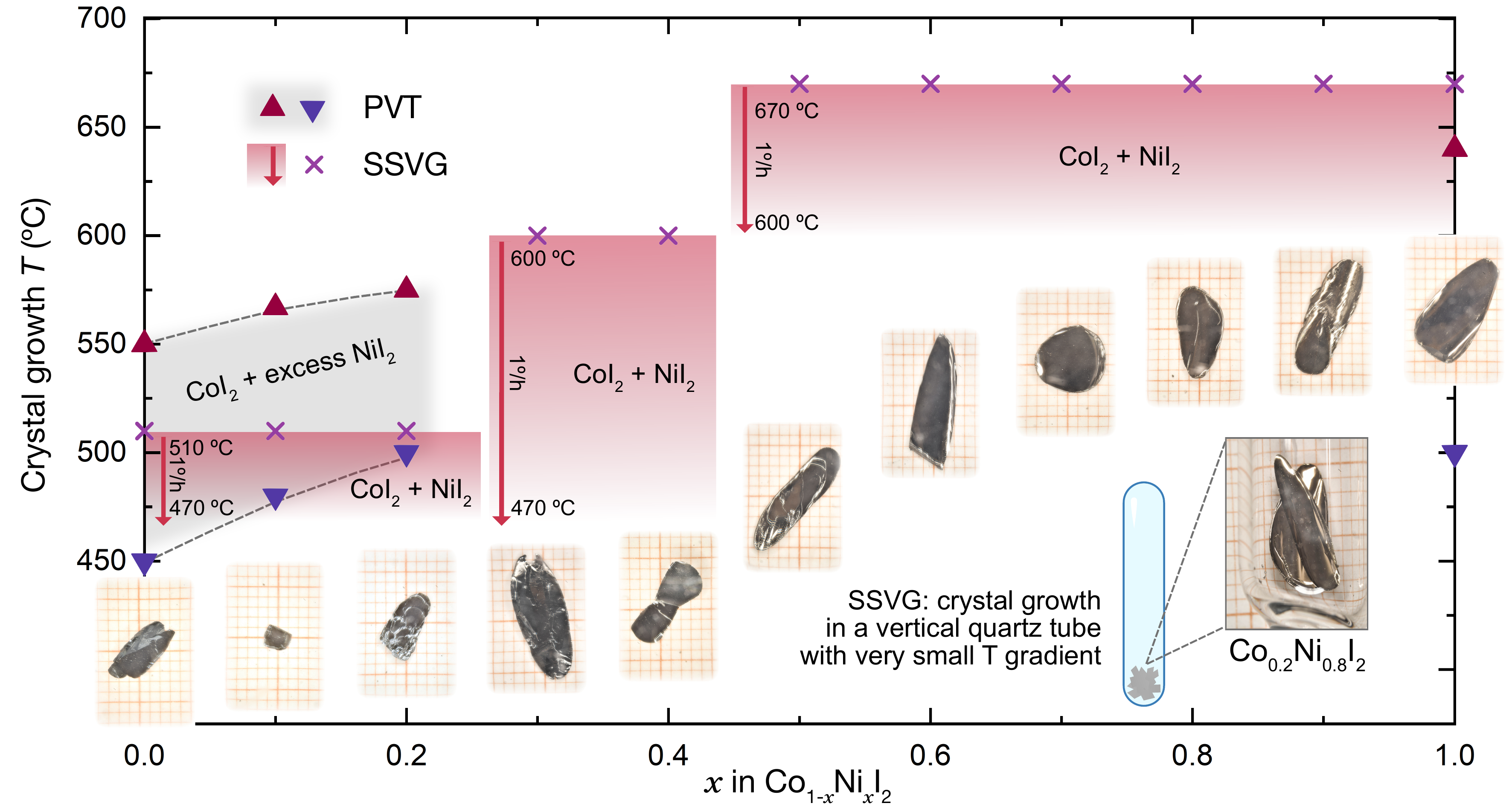}
    \caption{Crystal growth conditions of Co$_{1-x}$Ni$_x$I$_2$ by PVT and SSVG methods. Red and blue triangles show temperature gradients for PVT ($T_2 \to T_1$ respectively) for compounds with high Co content and pure CoI$_2$ and NiI$_2$. Red gradient rectangles represent temperatures and cooling rates for SSVG for the various compositions (purple crosses) of the solid solution. Pictures of crystals grown by SSVG were taken with an optical microscope on millimeter paper and correspond to the compositions on the $x$-axis. The inset shows an example of crystals as-grown by SSVG in a quartz tube.}
    \label{fig:CG}
\end{figure*}

PVT could also be used to grow crystals of Co$_{0.9}$Ni$_{0.1}$I$_2$ and Co$_{0.8}$Ni$_{0.2}$I$_2$. However, the obtained crystals with these stoichiometries were found to be smaller, with a maximal size of approximately $3.0\times3.0\times0.1$ mm$^3$. In this case the the PVT was found to be more challenging: the proper temperature gradient had to be adjusted precisely and a large (double) excess of NiI$_2$ had to be used to suppress the difference in the vapor pressures, as marked in the synthesis map in Figure \ref{fig:CG}. 

At a higher Ni content, PVT was found not to be a suitable method for crystal growth. With increasing the Ni content in Co$_{1-x}$Ni$_x$I$_2$, the reaction temperature increases as well, which causes the formation of crystals with intermediate and not homogeneous Co/Ni ratios. These limitations could not be overcome, despite careful adjustment of synthesis temperatures and the temperature gradients of the PVT synthesis. Naturally, these inhomogeneous crystals cannot be used for a detailed investigation of the physical properties of the solid solution. 

In the PVT growth experiments, we systematically observed that crystals could grow at both the cold and hot ends of the reactor. This observation suggested that another vapor-based growth process could be successful, namely the SSVG. This technique allows the avoidance of spatial separation of the starting constituents in the reactor and favors the formation of a compound with a precise composition. Here, we have developed the necessary synthesis conditions for realizing large single crystals of diiodides by SSVG, namely the whole Co$_{1-x}$Ni$_x$I$_2$ solid solution. Furthermore, the crystal size for the compositions with $x > 0.2 $ is only limited by the initial amount of employed chemicals and the diameter of the reaction vessel, as all of the starting material is converted into one or a few large crystals. The maximum size of these crystals obtained in our experiments was $6.0\times11.0\times0.4$ mm$^3$. It is noteworthy that we find that crystals grown by SSVG are thicker than those grown by PVT. The latter are usually rather thin flakes growing predominantly in the $(001)$ crystallographic plane. Grown crystals of pure CoI$_2$ and with low Ni content are shiny black/gray and those of pure NiI$_2$ and high Ni content are slightly goldish. The photographs of the crystals obtained by SSVG are shown in Figure \ref{fig:CG}.

In detail, we have adjusted the SSVG temperature ranges over three different regions of the compositions, namely $510-470$ \textdegree C ($0 \leq x < 0.3$), $600-470$ \textdegree C ($0.3 \leq x < 0.5$), and $670-600$ \textdegree C ($0.5 \leq x \leq 1$), and we applied a slow cooling procedure of 1 \textdegree C/h in every case. The details of the different temperature ranges, and the crystal compositions grown, are summarized in Figure \ref{fig:CG}. We observed that in the synthesis of Co-rich crystals, the partial melting of the precursors on the hot walls of the reactor can result in a phase separation, culminating in the formation of nearly pure CoI$_2$. To address this issue in the SSVG process, we initiated the reaction at a temperature below the melting point of pure CoI$_2$, which is $515$ \textdegree C, specifically for crystals with a Ni content of less than 0.3. 

We find that for this technique, the use of purified starting materials is crucial, as there is no spatial transport in a reactor. Furthermore, impurities already present in the starting materials may affect the vapor pressure during synthesis and therefore impact crystal growth as well. A cooling rate slower than 1 \textdegree C/h is not required, since the position in a furnace and the presence of impurities have a greater influence on the crystal quality than the cooling rate.

\subsection{Crystal Structure and Composition}

\begin{figure*}[h]
    \centering
    \includegraphics[width=\textwidth]{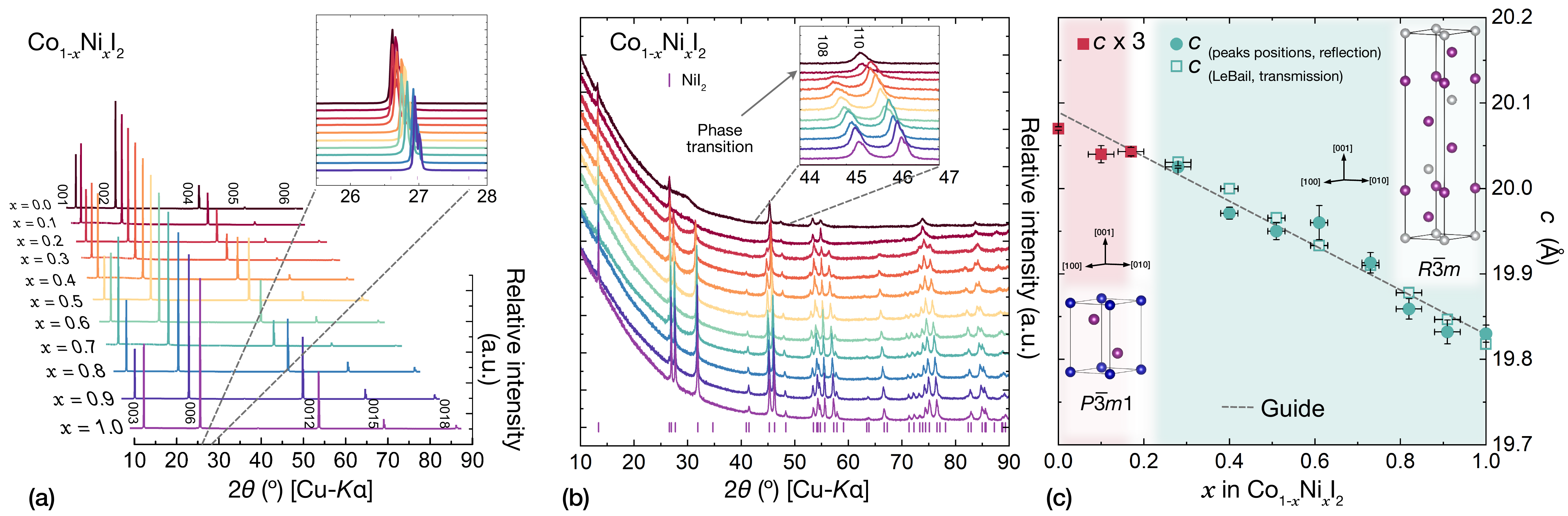}
    \caption{XRD analysis of the crystal structure of the solid solution Co$_{1-x}$Ni$_x$I$_2$. (a) XRD in the reflection Bragg-Brentano geometry on plate-like crystals laying parallel to the crystallographic $(001)$ plane. Only reflections from one family of planes are visible: $(00l)$, where $l=n$ for the CoI$_2$ side and $l=3n$ for the NiI$_2$ end. (b) PXRD in capillary mode for Co$_{1-x}$Ni$_x$I$_2$. In the inset, the peak $(110)$ characteristic for both crystal structures $P\bar{3}m1$ (CoI$_2$) and  $R\bar{3}m$ (NiI$_2$) is shifting along the solid solution. The (108) reflection is typical only for the NiI$_2$ structure and it vanishes completely for compositions with $x < 0.2$ indicating the phase transition. (c) The estimated cell parameter $c$ of solid solution Co$_{1-x}$Ni$_{x}$I$_2$ vs. composition. Solid green and red circles -- data obtained from the position of the peaks for the XRD in the reflection mode, open squares -- data from LeBail fitting for the capillaries measurements (error bars on $y$-axis smaller then points). For compounds with $P\Bar{3}m1$ crystal structure, the cell parameter is multiplied by 3 for clarity. Red and green colors represent regions for different crystal structures determined from SXRD and PXRD in capillary mode.}
    \label{fig:XRD}
\end{figure*}

Figure \ref{fig:XRD}(a) shows X-ray diffraction (XRD) patterns in the Bragg-Brentano geometry, i.e., reflection mode, on single plate-like crystals for all compositions of the Co$_{1-x}$Ni$_x$I$_2$ solid solution.

As discussed above, CoI$_2$ and NiI$_2$ crystallize both in the trigonal space groups $P\bar{3}m1$ and $R\bar{3}m$, respectively.\cite{Kuindersma1981} They have comparable $a=b$ cell parameters, and the difference between the two structures arises from the stacking of the layers (Figure \ref{fig:CoI2_NiI2}). The unit cell of NiI$_2$ is approximately three times larger and contains three layers and three formula units, as for CoI$_2$ there is only one layer and one formula unit per unit cell (Figure \ref{fig:XRD}(c)). The layers of these van der Waals materials lie parallel to the $(00l)$ planes. Henceforth, the XRD in the reflection mode on single crystals results in one family of reflections from $(00l)$ planes, where $l=n$ for the Co-rich side of solid solutions and $l=3n$ for the Ni-rich side. 

Based on the shift of the reflections, we have estimated the lattice parameter $c$ for all compositions. In Figure \ref{fig:XRD}(c), we show the linear trend of the $c$ parameter as a function of the Ni-content along the solid solution according to Vegard's law, the $c$ parameter of CoI$_2$ is multiplied by three as the unit cells of the end compounds differ by a factor of $3$.

The structural transition can not be tracked by the discontinuity of these reflections for different crystal structures, since these reflections belong to the same family of crystallographic planes and appear at the same 2$\theta$ angles. To determine the structural phase transition, we perform PXRD measurements (on finely grinded with starch crystals) in the Debye-Scherrer geometry, i.e., transmission mode, using filled capillaries. The PXRD pattern of all synthesized compositions are shown in Figure \ref{fig:XRD}(b). 

In these PXRD patterns, there are no visible impurity reflections, however, we observe a broadening and an absence of some $(h0l)$ reflections for CoI$_2$ and Co-rich compounds, which is probably due to mechanically-introduced disorder in the microstructure after grinding and stacking faults. This can be expected as the crystals are very soft and have weak van der Waals interactions between the layers. This effect was observed earlier by neutron diffraction in these systems of layered materials.\cite{Kuindersma1981} The change of crystal structure can be well observed from the PXRD patterns by tracking the evolution of the reflections of NiI$_2$ along the solid solution. For example, the $(108)$ peak typical only for NiI$_2$ structure (see Supplementary Information) shifts with the decrease of the Ni content and completely vanishes at the composition Co$_{0.9}$Ni$_{0.1}$I$_2$ (Figure \ref{fig:XRD}(b)). Thus, we find that the transition occurs around a composition of Co$_{0.8}$Ni$_{0.2}$I$_2$. Hence, only 20 \% of Ni atoms in CoI$_2$ switch the crystal structure from $P\bar{3}m1$ to $R\bar{3}m$ space group. This result is supported by our SXRD measurements as we found the $R\Bar{3}m$ crystal structures for compositions Co$_{0.2}$Ni$_{0.8}$I$_2$ and Co$_{0.7}$Ni$_{0.3}$I$_2$ (see Supplementary Information). From the PXRD data for compounds with the NiI$_2$ structure, using Le Bail fitting we obtained the lattice parameters $a=b$ (see Supplementary Information), and $c$ (Figure \ref{fig:XRD}(c)). The latest is consistent with the estimation from X-ray diffraction in the reflection mode. We did not perform Le Bail fitting for compounds with the CoI$_2$ structure due to the absence and broadening of some reflections, as discussed earlier.  

We find that with rising the Co content, the crystals of the solid solution are increasingly moisture sensitive. The SEM images (see Supplementary Information) show the appearance of bubbles and cracks on the crystals’ surface even after short exposure to air. These signs of crystal decomposition due to moisture interaction are visible for the solid solutions with a Ni content of less than 0.5 per formula unit, and their amount increases with the decreasing Ni substitution degree. 

\begin{figure} [t]
    \centering
    \includegraphics[width=3.33in]{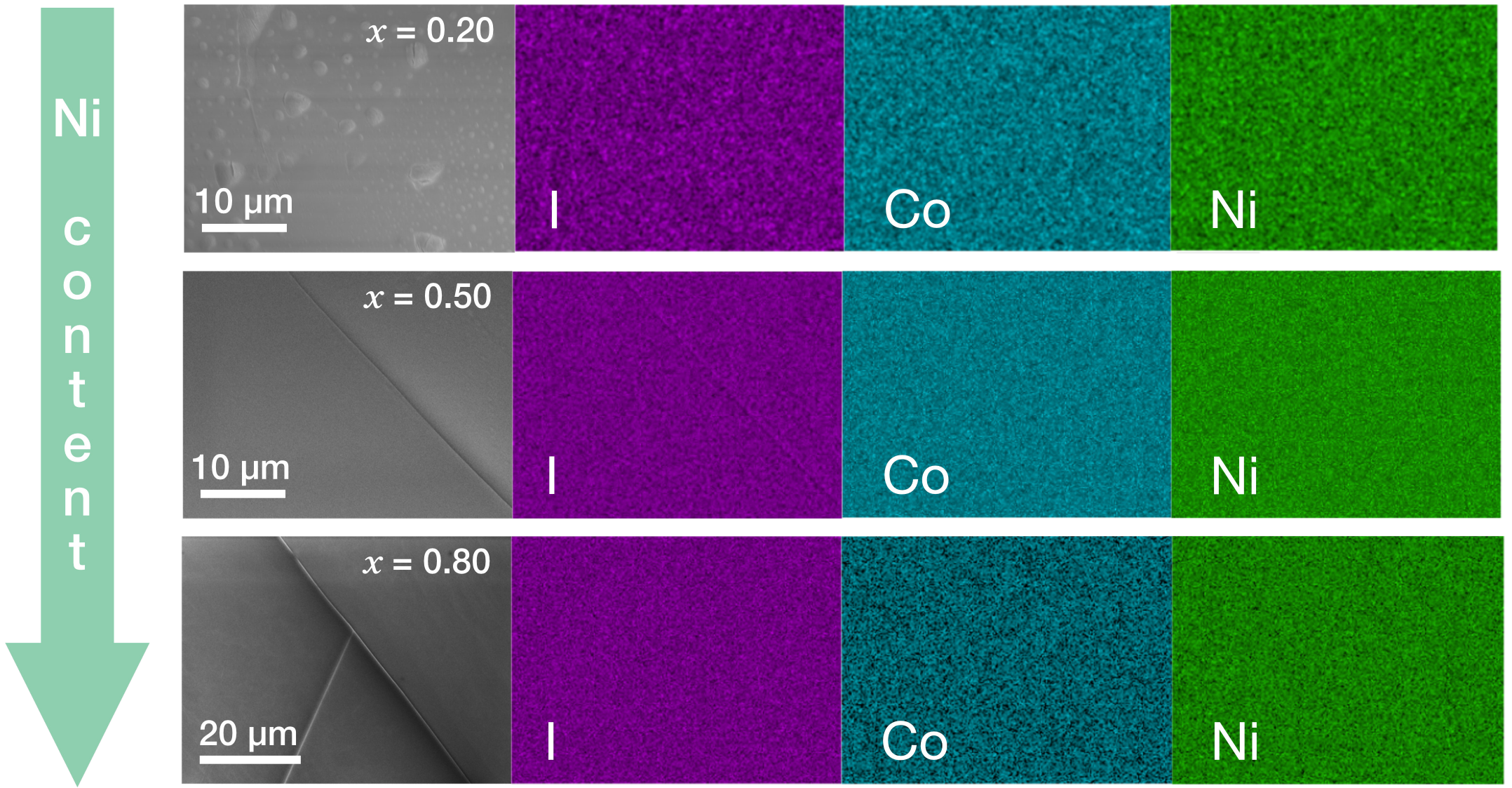}
    \caption{EDS analysis of crystals grown by SSVG from different ranges of the solid solution Co$_{1-x}$Ni$_x$I$_2$. For all compositions, Co and Ni atoms are distributed homogeneously without phase separation. Bubbles for the composition with $x = 0.2$ appeared on the surface due to crystal degradation on air during transfer to the SEM chamber.}
    \label{fig:EDX}
\end{figure}

EDS analysis confirms that the compositions of the crystals are homogeneous. In Figure \ref{fig:EDX}, the elemental distributions for three representative stoichiometries, namely $x$ = 0.2, 0.5, and 0.8 are presented. The uniform distribution of Co and Ni atoms is evidenced in the absence of any phase separation. The crystals grown with SSVG have the stoichiometrically expected composition, with an average error for Ni content around $\Delta x = \pm (0.01-0.02)$ (see Supplementary Information). In the PVT crystals, we find less uniform compositions and a larger average error around $\Delta x = \pm 0.03$. This compositional gradient for the crystals grown by PVT remains appropriate for physical measurements for Co-rich samples (up to $x \approx 0.2$ per formula unit). Nevertheless, with increasing Ni substitution degree, the gradient reaches tens of percent and PVT no longer becomes a suitable technique for synthesis of this solid solution, as discussed above. 

\subsection{Magnetic Properties}

\begin{figure*}
    \includegraphics[width=17cm]{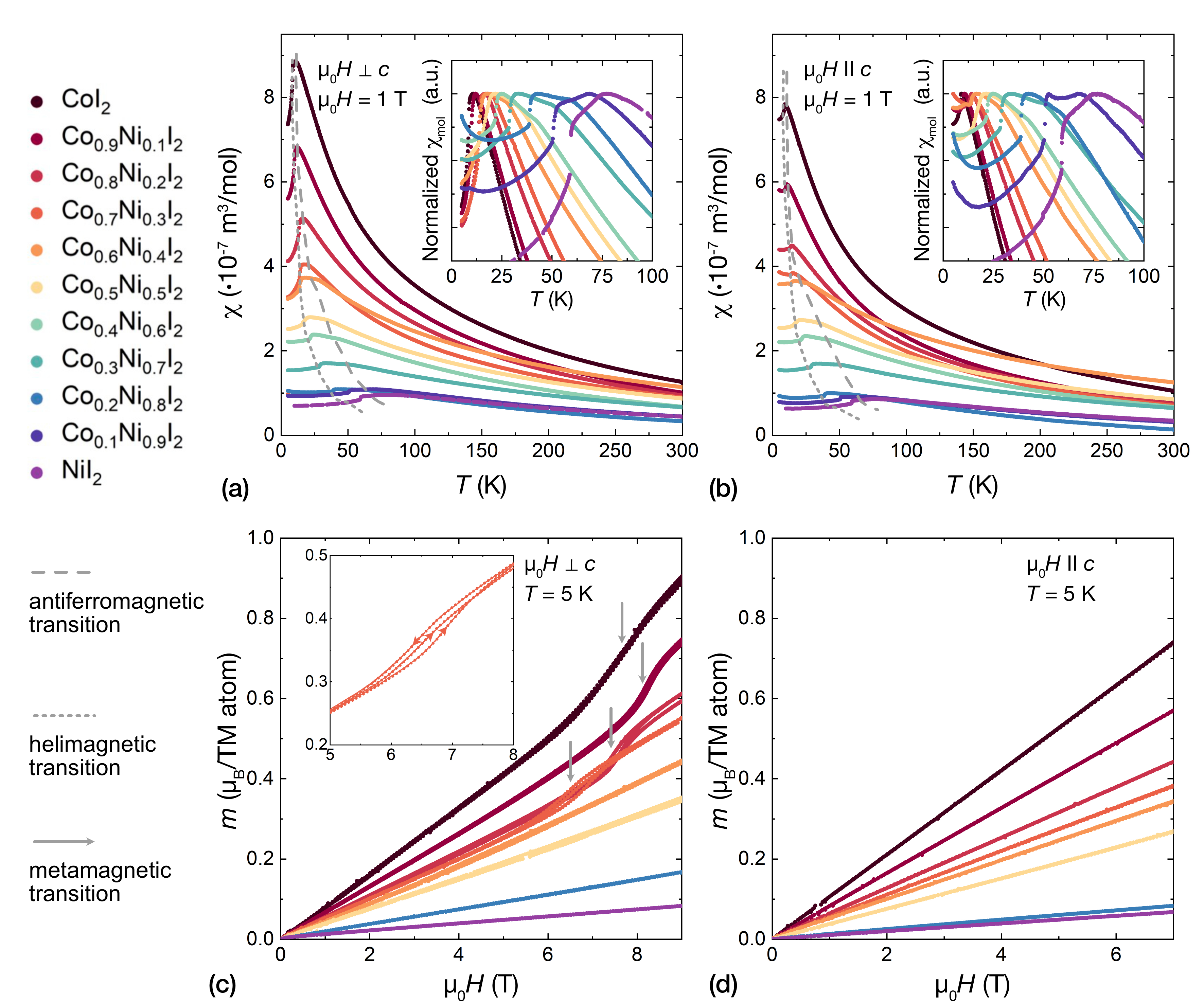}
    \caption{Anisotropic magnetic measurements of the solid solution Co$_{1-x}$Ni$_x$I$_2$. (a),(b) Magnetic susceptibility vs. temperature measurements parallel and perpendicular to the $c$-axis in a field of $\upmu_0 H = 1$ T. The dashed lines show the evolution of the first transition to the antiferromagnetic state ($T_{\rm N}$) and the dotted lines show the second transition to helimagnetic order ($T_{\rm HM}$). Insets show a zoom in of the normalized magnetic susceptibility, displaying the two magnetic transitions. (c),(d) Measurements of magnetic moment vs. magnetic field parallel and perpendicular to the $c$-axis at $T$ = 5 K. A gray arrow marks the deviation from linearity and the hysteresis, which is associated with the metamagnetic transition known for CoI$_2$. The inset in (c) shows an enlarged hysteresis for Co$_{0.7}$Ni$_{0.3}$I$_2$. The orange arrows indicate the direction of change of the applied magnetic field.}
    \label{fig.MT_MH}
\end{figure*}

We studied the anisotropic magnetic properties of the solid solution by measuring the temperature and field dependence of the magnetic moment, in a magnetic field applied either parallel or perpendicular to the crystallographic $c$-axis.
In Figure \ref{fig.MT_MH}(a) \& (b), we present the temperature-dependent molar magnetic susceptibility for all synthesized members of the solid solution Co$_{1-x}$Ni$_x$I$_2$. At high temperatures, we observe a Curie-Weiss-like behavior for all compositions, which agrees well with the presence of a paramagnetic state, corresponding to the localized electrons in the $d$-orbitals of the transition metals. 

At lower temperatures, all samples display transitions to long-range magnetically ordered states. Specifically, we observe a transition at the Néel temperature $T_{\rm N}$ to an antiferromagnetically ordered state, which corresponds to a first kink in the magnetic susceptibility data. At even lower temperatures ($T_{\rm HM}$), we identified the transition to the helimagnetic state as the sudden jump in the magnetic susceptibility. 

The respective transition temperatures that we observe for the end members of our solid solution are in excellent agreement with earlier reports.\cite{Billerey1977, Billerey1980, Mekata1992, Kurumaji2013} We find that both transition temperatures change smoothly and continuously as a function of chemical composition from one end member to the other. 

At low temperatures (below all the transitions, in the magnetic ground state) the magnetic response of the Co-rich samples exhibits a clear dependence on the direction of the applied magnetic field (Figure 5(a) \& (b)). This behavior is not visible on the Ni-rich side of the solid solution, in which the helical magnetic order is different. 

Field-dependent anisotropic measurements of the magnetic moment are presented in Figure \ref{fig.MT_MH}(c) \& (d). For both crystal orientations, we observe no saturation of the magnetic moment for any compositions of the solid solution up to the applied magnetic field of $\upmu_0 H = 9 T$. We find magnetic moments that are not higher than 1 $\upmu_B$ per transition metal ion. Even in the high fields, they do not reach the theoretically possible values that are expected for spin-magnetism in an octahedral ligand field: $\mu_{\rm calc.} = 2\upmu_{\rm B}$ for NiI$_2$ ($S = 1$) and $\mu_{\rm calc.} = 3\upmu_{\rm B}$ for CoI$_2$ (high-spin state, $S = \frac{3}{2}$). This discrepancy is due to the frustrated  helimagnetic order and the absence of a spin-flip out of the easy plane in this range of applied magnetic fields. 

When the magnetic field is applied parallel to the easy $(001)$-plane a distinct deviation from the linear slope and a hysteresis are observed in magnetic fields between $\upmu_0 H$ = 6 T and 9 T, for the Co-rich compounds with $0 \leq x \leq 0.3$ (Figure \ref{fig.MT_MH}(c)). This feature in magnetic moment $m$ corresponds to a metamagnetic phase transition, which had been reported for CoI$_2$ previously.\cite{Kurumaji2013} We observe that, as the Ni content increases, there is a noticeable shift and alteration in the hysteresis shape. Furthermore, when the Ni concentration exceeds 0.3 per formula unit, this transition phenomenon disappears entirely. Thus, a small Ni substitution for Co in CoI$_2$ (up to 30 $\%$) affects this metamagnetic transition by changing the temperature/field range of existence of this metamagnetic state. These properties can be controlled by the composition. 

\subsection{Magnetic Phase Diagram}

\begin{figure}
    \centering
    \includegraphics[width=17cm]{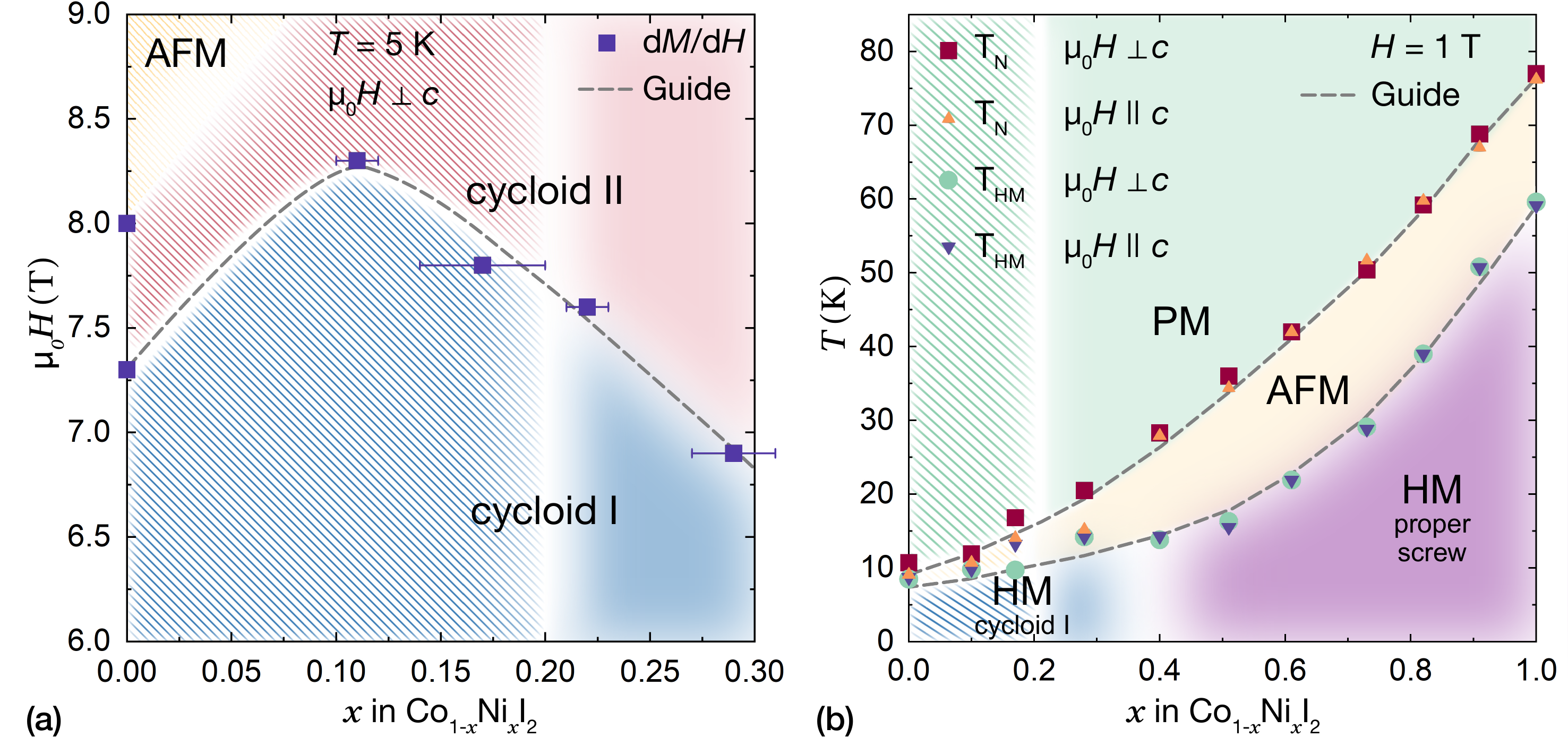}
    \caption{Phase diagrams of the magnetic properties of the Co$_{1-x}$Ni$_x$I$_2$ solid solution. The hatching background corresponds to the CoI$_2$ crystal structure, while the bold background represents the NiI$_2$ crystal structure at room temperature. (a) Magnetic field-composition phase diagram for the compositional range $0 \leq x \leq 0.3$. The blue background depicts the helimagnetic ground state, named cycloid I. The red background represents another type of spin arrangement, named cycloid II. The yellow background corresponds to a antiferromagnetic (AFM) state. The dashed line based on the magnetic field derivative of the magnetic moment (purple squares) shows the metamagnetic transition from one cycloid to the other. (b) Temperature-composition phase diagram for the whole range of the solid solution at low magnetic fields (i.e. $\upmu_0 H$ = 1 T). PM on the green background -- paramagnetic state, AFM on the purple background -- antiferromagnetic state, HM on the purple background -- proper screw helimagnetic order, HM on the blue background -- cycloid helimagnetic order.} 
    \label{fig:PD}
\end{figure}

In Figure \ref{fig:PD}(a), we present the magnetic field -- composition phase diagram for the compositional range $0 \leq x \leq 0.3$ at 5 K. This phase diagram was derived from the analysis of the maxima of the first derivative of the magnetic moment vs. magnetic field data ($\mathrm{d} m/\mathrm{d} H$) at high magnetic fields $\upmu_0 H > 6$ T (see Supplementary Information). The blue area corresponds to the helimagnetic ground state typical for CoI$_2$, denoted cycloid I. When the magnetic field overcomes a certain critical field strength, the magnetic order switches to another type of spiral order, denoted as cycloid II (red area). In this phase, the spins are expected to reorient within the plane perpendicular to the applied magnetic field according to earlier results on CoI$_2$.\cite{Kurumaji2013} For pure CoI$_2$ we observed the second maximum of the first derivative at the magnetic field around 8 T. This corresponds to the transition from cycloidal order to antiferromagnetic state which was also reported earlier for this compound.\cite{Kurumaji2013}. We did not observe this second transition for other members of the solid solution since it should occur at a higher magnetic field out of the range of our measurements.  As the structural phase transition in this system occurs around a composition of Co$_{0.8}$Ni$_{0.2}$I$_2$, the phase diagram may be divided into two regions: with crystal structures of CoI$_2$ (hatching background) and NiI$_2$ (solid background). Thus, we find the existence of a range of compositions of Co$_{1-x}$Ni$_x$I$_2$ ($0.2 \leq x \leq 0.3$), which host magnetic order with the metamagnetic transition similar to that for CoI$_2$ in the NiI$_2$ crystal structure. 

Based on the temperature-dependent magnetic susceptibility data and the field-dependent magnetic measurements of the Co$_{1-x}$Ni$_x$I$_2$, we are able to construct a comprehensive magnetic phase diagram, as presented in Figure \ref{fig:PD}(b). We have determined the transition temperatures, $T_{\rm N}$ and $T_{\rm HM}$ from the first derivative $\mathrm{d}\chi/\mathrm{d}T$ of magnetic susceptibility vs. temperature. We found almost identical transition temperatures for the measurements performed at different orientations of the applied magnetic fields.  

In the phase diagram, four different magnetic phases can be identified. The first is the paramagnetic (PM) state at high temperatures. The second one is the antiferromagnetic (AFM) state, which is located between lines corresponding to the $T_{\rm N}$ and $T_{\rm HM}$. The third and fourth phases are the two different helimagnetic (HM) ground states at low temperatures (below $T_{\rm HM}$ line). This we can conclude based on the observation of the metamagnetic transitions of the solid solution with Ni content below 0.3 per formula unit. It provides evidence that there are at least two different helimagnetic states present within the Co$_{1-x}$Ni$_x$I$_2$ solid solution: the proper screw order on the Ni-rich side and the cycloid state on the Co-rich side. The transition between these two magnetic orders occurs in the composition range $0.3 < x < 0.4$ since the metamagnetic transition vanishes for compounds with $x > 0.3$.  It is worth noticing that there is a discrepancy between the change of helical order (determined from the existence of the metamagnetic transition) and the structural transition along the solid solution. It means that for the intermediate compounds (in between the structural and the magnetic transitions) new spin states may emerge and host different magnetoelectric coupling. However, the exact  determination of this phase boundary requires further investigations. Nevertheless, we can conclude that all magnetic properties are continuously changing along the solid solution. This opens the possibility of tuning magnetic properties and creating a certain spin order in this system by adjusting the composition of a crystal.  

\section{Conclusions}
We report on the successful growth of crystals of the entire van der Waals magnet Co$_{1-x}$Ni$_x$I$_2$ solid solution and on its magnetic properties. We have successfully established the synthesis conditions to produce crystals of the Co$_{1-x}$Ni$_x$I$_2$ solid solution using SSVG. This method ensures the growth of crystals with uniform composition, as it prevents the spatial separation of initial substances in the reactor. Contrary to PVT, SSVG of these diiodides results in the formation of compounds with accurate compositions. Additionally, the size of the crystals for the solid solutions with Ni content $x > 0.2$ is determined solely by the quantity of the starting material used and the size of the synthesis vessel, as the entire starting material is transformed into a few large single crystals.

The lattice parameters of the solid solution Co$_{1-x}$Ni$_x$I$_2$ show a linear trend along all the compositions in accordance to the Vegard’s law. The magnetic order smoothly evolves in a non-linear fashion from one end member to the other. In the magnetic phase diagram, we identify four distinct magnetic states. The first is the paramagnetic (PM) state , which occurs at high temperatures. The second is the antiferromagnetic (AFM) state. The third and fourth states are two types of helimagnetic ground states at low temperatures, occurring below $T_{\rm HM}$. We have found that Ni substitution in CoI$_2$ also affects metamagnetic transition at high fields and allows control of the temperature-field range of this transition. 
Bulk crystals of the solid solution Co$_{1-x}$Ni$_x$I$_2$ hold a possibility to be exfoliated into 2D thin flakes and are expected to have thickness-dependent magnetic behavior down to a monolayer, thus opening the possibility of fabricating new 2D magnetic devices with novel magnetic properties.


\begin{acknowledgement}

This work was supported  by the Swiss National Science Foundation under Grants No. PCEFP2\_194183 and No. 200021-204065. The authors would like to thank Dmitry Lebedev for valuable discussions, Siobhan McKeown Walker for help with SEM and EDS measurements, and the entire von Rohr group for practical advice and support. 

\end{acknowledgement}

\begin{suppinfo}

The Supporting Information is available free of charge online.

SEM images of cleaved crystals, targeting compositions and obtained from EDS analysis, SXRD data, Le Bail fitting of the PXRD data, evolution of $a = b$ cell parameters, simulated powder patterns, magnetic measurements up to 9 T magnetic field applied parallel to $c$-axis, $\mathrm{d}m/\mathrm{d}H$ derivatives of the magnetic moment vs. magnetic field data. 

\end{suppinfo}

\bibliography{Co1-xNixI2_v3_ACS}

\providecommand{\latin}[1]{#1}
\makeatletter
\providecommand{\doi}
  {\begingroup\let\do\@makeother\dospecials
  \catcode`\{=1 \catcode`\}=2 \doi@aux}
\providecommand{\doi@aux}[1]{\endgroup\texttt{#1}}
\makeatother
\providecommand*\mcitethebibliography{\thebibliography}
\csname @ifundefined\endcsname{endmcitethebibliography}  {\let\endmcitethebibliography\endthebibliography}{}
\begin{mcitethebibliography}{39}
\providecommand*\natexlab[1]{#1}
\providecommand*\mciteSetBstSublistMode[1]{}
\providecommand*\mciteSetBstMaxWidthForm[2]{}
\providecommand*\mciteBstWouldAddEndPuncttrue
  {\def\EndOfBibitem{\unskip.}}
\providecommand*\mciteBstWouldAddEndPunctfalse
  {\let\EndOfBibitem\relax}
\providecommand*\mciteSetBstMidEndSepPunct[3]{}
\providecommand*\mciteSetBstSublistLabelBeginEnd[3]{}
\providecommand*\EndOfBibitem{}
\mciteSetBstSublistMode{f}
\mciteSetBstMaxWidthForm{subitem}{(\alph{mcitesubitemcount})}
\mciteSetBstSublistLabelBeginEnd
  {\mcitemaxwidthsubitemform\space}
  {\relax}
  {\relax}

\bibitem[Novoselov \latin{et~al.}(2004)Novoselov, Geim, Morozov, Jiang, Zhang, Dubonos, Grigorieva, and Firsov]{novoselov2004electric}
Novoselov,~K.~S.; Geim,~A.~K.; Morozov,~S.~V.; Jiang,~D.-e.; Zhang,~Y.; Dubonos,~S.~V.; Grigorieva,~I.~V.; Firsov,~A.~A. Electric field effect in atomically thin carbon films. \emph{Science} \textbf{2004}, \emph{306}, 666--669\relax
\mciteBstWouldAddEndPuncttrue
\mciteSetBstMidEndSepPunct{\mcitedefaultmidpunct}
{\mcitedefaultendpunct}{\mcitedefaultseppunct}\relax
\EndOfBibitem
\bibitem[Whittingham(1978)]{whittingham1978chemistry}
Whittingham,~M.~S. Chemistry of intercalation compounds: Metal guests in chalcogenide hosts. \emph{Progress in Solid State Chemistry} \textbf{1978}, \emph{12}, 41--99\relax
\mciteBstWouldAddEndPuncttrue
\mciteSetBstMidEndSepPunct{\mcitedefaultmidpunct}
{\mcitedefaultendpunct}{\mcitedefaultseppunct}\relax
\EndOfBibitem
\bibitem[Hulliger(2012)]{hulliger2012structural}
Hulliger,~F. \emph{Structural chemistry of layer-type phases}; Springer Science \& Business Media, 2012; Vol.~5\relax
\mciteBstWouldAddEndPuncttrue
\mciteSetBstMidEndSepPunct{\mcitedefaultmidpunct}
{\mcitedefaultendpunct}{\mcitedefaultseppunct}\relax
\EndOfBibitem
\bibitem[Manzeli \latin{et~al.}(2017)Manzeli, Ovchinnikov, Pasquier, Yazyev, and Kis]{manzeli20172d}
Manzeli,~S.; Ovchinnikov,~D.; Pasquier,~D.; Yazyev,~O.~V.; Kis,~A. 2D transition metal dichalcogenides. \emph{Nature Reviews Materials} \textbf{2017}, \emph{2}, 1--15\relax
\mciteBstWouldAddEndPuncttrue
\mciteSetBstMidEndSepPunct{\mcitedefaultmidpunct}
{\mcitedefaultendpunct}{\mcitedefaultseppunct}\relax
\EndOfBibitem
\bibitem[Zheng \latin{et~al.}(2021)Zheng, Wilfong, Hickox-Young, Rondinelli, Zavalij, and Rodriguez]{zheng2021polar}
Zheng,~H.; Wilfong,~B.~C.; Hickox-Young,~D.; Rondinelli,~J.~M.; Zavalij,~P.~Y.; Rodriguez,~E.~E. Polar ferromagnetic metal by intercalation of metal--amine complexes. \emph{Chemistry of Materials} \textbf{2021}, \emph{33}, 4936--4947\relax
\mciteBstWouldAddEndPuncttrue
\mciteSetBstMidEndSepPunct{\mcitedefaultmidpunct}
{\mcitedefaultendpunct}{\mcitedefaultseppunct}\relax
\EndOfBibitem
\bibitem[Geim and Grigorieva(2013)Geim, and Grigorieva]{geim2013van}
Geim,~A.~K.; Grigorieva,~I.~V. Van der Waals heterostructures. \emph{Nature} \textbf{2013}, \emph{499}, 419--425\relax
\mciteBstWouldAddEndPuncttrue
\mciteSetBstMidEndSepPunct{\mcitedefaultmidpunct}
{\mcitedefaultendpunct}{\mcitedefaultseppunct}\relax
\EndOfBibitem
\bibitem[Duong \latin{et~al.}(2017)Duong, Yun, and Lee]{duong2017van}
Duong,~D.~L.; Yun,~S.~J.; Lee,~Y.~H. van der Waals layered materials: opportunities and challenges. \emph{ACS nano} \textbf{2017}, \emph{11}, 11803--11830\relax
\mciteBstWouldAddEndPuncttrue
\mciteSetBstMidEndSepPunct{\mcitedefaultmidpunct}
{\mcitedefaultendpunct}{\mcitedefaultseppunct}\relax
\EndOfBibitem
\bibitem[Gong \latin{et~al.}(2017)Gong, Li, Li, Ji, Stern, Xia, Cao, Bao, Wang, Wang, \latin{et~al.} others]{gong2017discovery}
Gong,~C.; Li,~L.; Li,~Z.; Ji,~H.; Stern,~A.; Xia,~Y.; Cao,~T.; Bao,~W.; Wang,~C.; Wang,~Y.; others Discovery of intrinsic ferromagnetism in two-dimensional van der Waals crystals. \emph{Nature} \textbf{2017}, \emph{546}, 265--269\relax
\mciteBstWouldAddEndPuncttrue
\mciteSetBstMidEndSepPunct{\mcitedefaultmidpunct}
{\mcitedefaultendpunct}{\mcitedefaultseppunct}\relax
\EndOfBibitem
\bibitem[Huang \latin{et~al.}(2017)Huang, Clark, Navarro-Moratalla, Klein, Cheng, Seyler, Zhong, Schmidgall, McGuire, Cobden, \latin{et~al.} others]{huang2017layer}
Huang,~B.; Clark,~G.; Navarro-Moratalla,~E.; Klein,~D.~R.; Cheng,~R.; Seyler,~K.~L.; Zhong,~D.; Schmidgall,~E.; McGuire,~M.~A.; Cobden,~D.~H.; others Layer-dependent ferromagnetism in a van der Waals crystal down to the monolayer limit. \emph{Nature} \textbf{2017}, \emph{546}, 270--273\relax
\mciteBstWouldAddEndPuncttrue
\mciteSetBstMidEndSepPunct{\mcitedefaultmidpunct}
{\mcitedefaultendpunct}{\mcitedefaultseppunct}\relax
\EndOfBibitem
\bibitem[Burch \latin{et~al.}(2018)Burch, Mandrus, and Park]{burch2018magnetism}
Burch,~K.~S.; Mandrus,~D.; Park,~J.-G. Magnetism in two-dimensional van der Waals materials. \emph{Nature} \textbf{2018}, \emph{563}, 47--52\relax
\mciteBstWouldAddEndPuncttrue
\mciteSetBstMidEndSepPunct{\mcitedefaultmidpunct}
{\mcitedefaultendpunct}{\mcitedefaultseppunct}\relax
\EndOfBibitem
\bibitem[Gibertini \latin{et~al.}(2019)Gibertini, Koperski, Morpurgo, and Novoselov]{gibertini2019magnetic}
Gibertini,~M.; Koperski,~M.; Morpurgo,~A.~F.; Novoselov,~K.~S. Magnetic 2D materials and heterostructures. \emph{Nature nanotechnology} \textbf{2019}, \emph{14}, 408--419\relax
\mciteBstWouldAddEndPuncttrue
\mciteSetBstMidEndSepPunct{\mcitedefaultmidpunct}
{\mcitedefaultendpunct}{\mcitedefaultseppunct}\relax
\EndOfBibitem
\bibitem[Sierra \latin{et~al.}(2021)Sierra, Fabian, Kawakami, Roche, and Valenzuela]{sierra2021van}
Sierra,~J.~F.; Fabian,~J.; Kawakami,~R.~K.; Roche,~S.; Valenzuela,~S.~O. Van der Waals heterostructures for spintronics and opto-spintronics. \emph{Nature Nanotechnology} \textbf{2021}, \emph{16}, 856--868\relax
\mciteBstWouldAddEndPuncttrue
\mciteSetBstMidEndSepPunct{\mcitedefaultmidpunct}
{\mcitedefaultendpunct}{\mcitedefaultseppunct}\relax
\EndOfBibitem
\bibitem[Song \latin{et~al.}(2018)Song, Cai, Tu, Zhang, Huang, Wilson, Seyler, Zhu, Taniguchi, Watanabe, \latin{et~al.} others]{song2018giant}
Song,~T.; Cai,~X.; Tu,~M. W.-Y.; Zhang,~X.; Huang,~B.; Wilson,~N.~P.; Seyler,~K.~L.; Zhu,~L.; Taniguchi,~T.; Watanabe,~K.; others Giant tunneling magnetoresistance in spin-filter van der Waals heterostructures. \emph{Science} \textbf{2018}, \emph{360}, 1214--1218\relax
\mciteBstWouldAddEndPuncttrue
\mciteSetBstMidEndSepPunct{\mcitedefaultmidpunct}
{\mcitedefaultendpunct}{\mcitedefaultseppunct}\relax
\EndOfBibitem
\bibitem[N{\v{e}}mec \latin{et~al.}(2018)N{\v{e}}mec, Fiebig, Kampfrath, and Kimel]{nvemec2018antiferromagnetic}
N{\v{e}}mec,~P.; Fiebig,~M.; Kampfrath,~T.; Kimel,~A.~V. Antiferromagnetic opto-spintronics. \emph{Nature Physics} \textbf{2018}, \emph{14}, 229--241\relax
\mciteBstWouldAddEndPuncttrue
\mciteSetBstMidEndSepPunct{\mcitedefaultmidpunct}
{\mcitedefaultendpunct}{\mcitedefaultseppunct}\relax
\EndOfBibitem
\bibitem[Wang \latin{et~al.}(2019)Wang, Gibertini, Dumcenco, Taniguchi, Watanabe, Giannini, and Morpurgo]{wang2019determining}
Wang,~Z.; Gibertini,~M.; Dumcenco,~D.; Taniguchi,~T.; Watanabe,~K.; Giannini,~E.; Morpurgo,~A.~F. Determining the phase diagram of atomically thin layered antiferromagnet CrCl$_3$. \emph{Nature nanotechnology} \textbf{2019}, \emph{14}, 1116--1122\relax
\mciteBstWouldAddEndPuncttrue
\mciteSetBstMidEndSepPunct{\mcitedefaultmidpunct}
{\mcitedefaultendpunct}{\mcitedefaultseppunct}\relax
\EndOfBibitem
\bibitem[Bonilla \latin{et~al.}(2018)Bonilla, Kolekar, Ma, Diaz, Kalappattil, Das, Eggers, Gutierrez, Phan, and Batzill]{bonilla2018strong}
Bonilla,~M.; Kolekar,~S.; Ma,~Y.; Diaz,~H.~C.; Kalappattil,~V.; Das,~R.; Eggers,~T.; Gutierrez,~H.~R.; Phan,~M.-H.; Batzill,~M. Strong room-temperature ferromagnetism in VSe$_2$ monolayers on van der Waals substrates. \emph{Nature nanotechnology} \textbf{2018}, \emph{13}, 289--293\relax
\mciteBstWouldAddEndPuncttrue
\mciteSetBstMidEndSepPunct{\mcitedefaultmidpunct}
{\mcitedefaultendpunct}{\mcitedefaultseppunct}\relax
\EndOfBibitem
\bibitem[Wu \latin{et~al.}(2023)Wu, Gibertini, Watanabe, Taniguchi, Guti{\'e}rrez-Lezama, Ubrig, and Morpurgo]{wu2023gate}
Wu,~F.; Gibertini,~M.; Watanabe,~K.; Taniguchi,~T.; Guti{\'e}rrez-Lezama,~I.; Ubrig,~N.; Morpurgo,~A.~F. Gate-Controlled Magnetotransport and Electrostatic Modulation of Magnetism in 2D Magnetic Semiconductor CrPS$_4$. \emph{Advanced Materials} \textbf{2023}, 2211653\relax
\mciteBstWouldAddEndPuncttrue
\mciteSetBstMidEndSepPunct{\mcitedefaultmidpunct}
{\mcitedefaultendpunct}{\mcitedefaultseppunct}\relax
\EndOfBibitem
\bibitem[Telford \latin{et~al.}(2020)Telford, Dismukes, Lee, Cheng, Wieteska, Bartholomew, Chen, Xu, Pasupathy, Zhu, \latin{et~al.} others]{telford2020layered}
Telford,~E.~J.; Dismukes,~A.~H.; Lee,~K.; Cheng,~M.; Wieteska,~A.; Bartholomew,~A.~K.; Chen,~Y.-S.; Xu,~X.; Pasupathy,~A.~N.; Zhu,~X.; others Layered antiferromagnetism induces large negative magnetoresistance in the van der Waals semiconductor CrSBr. \emph{Advanced Materials} \textbf{2020}, \emph{32}, 2003240\relax
\mciteBstWouldAddEndPuncttrue
\mciteSetBstMidEndSepPunct{\mcitedefaultmidpunct}
{\mcitedefaultendpunct}{\mcitedefaultseppunct}\relax
\EndOfBibitem
\bibitem[Wilson \latin{et~al.}(2021)Wilson, Lee, Cenker, Xie, Dismukes, Telford, Fonseca, Sivakumar, Dean, Cao, \latin{et~al.} others]{wilson2021interlayer}
Wilson,~N.~P.; Lee,~K.; Cenker,~J.; Xie,~K.; Dismukes,~A.~H.; Telford,~E.~J.; Fonseca,~J.; Sivakumar,~S.; Dean,~C.; Cao,~T.; others Interlayer electronic coupling on demand in a 2D magnetic semiconductor. \emph{Nature Materials} \textbf{2021}, \emph{20}, 1657--1662\relax
\mciteBstWouldAddEndPuncttrue
\mciteSetBstMidEndSepPunct{\mcitedefaultmidpunct}
{\mcitedefaultendpunct}{\mcitedefaultseppunct}\relax
\EndOfBibitem
\bibitem[Wu \latin{et~al.}(2022)Wu, Guti{\'e}rrez-Lezama, L{\'o}pez-Paz, Gibertini, Watanabe, Taniguchi, von Rohr, Ubrig, and Morpurgo]{wu2022quasi}
Wu,~F.; Guti{\'e}rrez-Lezama,~I.; L{\'o}pez-Paz,~S.~A.; Gibertini,~M.; Watanabe,~K.; Taniguchi,~T.; von Rohr,~F.~O.; Ubrig,~N.; Morpurgo,~A.~F. Quasi-1D Electronic Transport in a 2D Magnetic Semiconductor. \emph{Advanced Materials} \textbf{2022}, \emph{34}, 2109759\relax
\mciteBstWouldAddEndPuncttrue
\mciteSetBstMidEndSepPunct{\mcitedefaultmidpunct}
{\mcitedefaultendpunct}{\mcitedefaultseppunct}\relax
\EndOfBibitem
\bibitem[L{\'o}pez-Paz \latin{et~al.}(2022)L{\'o}pez-Paz, Guguchia, Pomjakushin, Witteveen, Cervellino, Luetkens, Casati, Morpurgo, and von Rohr]{lopez2022dynamic}
L{\'o}pez-Paz,~S.~A.; Guguchia,~Z.; Pomjakushin,~V.~Y.; Witteveen,~C.; Cervellino,~A.; Luetkens,~H.; Casati,~N.; Morpurgo,~A.~F.; von Rohr,~F.~O. Dynamic magnetic crossover at the origin of the hidden-order in van der Waals antiferromagnet CrSBr. \emph{Nature Communications} \textbf{2022}, \emph{13}, 4745\relax
\mciteBstWouldAddEndPuncttrue
\mciteSetBstMidEndSepPunct{\mcitedefaultmidpunct}
{\mcitedefaultendpunct}{\mcitedefaultseppunct}\relax
\EndOfBibitem
\bibitem[McGuire(2017)]{McGuire2017}
McGuire,~M.~A. Crystal and magnetic structures in layered, transition metal dihalides and trihalides. \emph{Crystals} \textbf{2017}, \emph{7}, 121\relax
\mciteBstWouldAddEndPuncttrue
\mciteSetBstMidEndSepPunct{\mcitedefaultmidpunct}
{\mcitedefaultendpunct}{\mcitedefaultseppunct}\relax
\EndOfBibitem
\bibitem[Khomskii(2009)]{Khomskii2009}
Khomskii,~D. Classifying multiferroics: Mechanisms and effects. \emph{Physics} \textbf{2009}, \emph{2}, 20\relax
\mciteBstWouldAddEndPuncttrue
\mciteSetBstMidEndSepPunct{\mcitedefaultmidpunct}
{\mcitedefaultendpunct}{\mcitedefaultseppunct}\relax
\EndOfBibitem
\bibitem[Tokura and Seki(2010)Tokura, and Seki]{Tokura2010}
Tokura,~Y.; Seki,~S. Multiferroics with spiral spin orders. \emph{Advanced materials} \textbf{2010}, \emph{22}, 1554--1565\relax
\mciteBstWouldAddEndPuncttrue
\mciteSetBstMidEndSepPunct{\mcitedefaultmidpunct}
{\mcitedefaultendpunct}{\mcitedefaultseppunct}\relax
\EndOfBibitem
\bibitem[Son \latin{et~al.}(2022)Son, Lee, Kim, Kim, Kim, Na, Ju, Park, Nag, Zhou, \latin{et~al.} others]{Son2022}
Son,~S.; Lee,~Y.; Kim,~J.~H.; Kim,~B.~H.; Kim,~C.; Na,~W.; Ju,~H.; Park,~S.; Nag,~A.; Zhou,~K.-J.; others Multiferroic-Enabled Magnetic-Excitons in 2D Quantum-Entangled Van der Waals Antiferromagnet NiI$_2$. \emph{Advanced Materials} \textbf{2022}, \emph{34}, 2109144\relax
\mciteBstWouldAddEndPuncttrue
\mciteSetBstMidEndSepPunct{\mcitedefaultmidpunct}
{\mcitedefaultendpunct}{\mcitedefaultseppunct}\relax
\EndOfBibitem
\bibitem[Lebedev \latin{et~al.}(2023)Lebedev, Gish, Garvey, Stanev, Choi, Georgopoulos, Song, Park, Watanabe, Taniguchi, \latin{et~al.} others]{Lebedev2023}
Lebedev,~D.; Gish,~J.~T.; Garvey,~E.~S.; Stanev,~T.~K.; Choi,~J.; Georgopoulos,~L.; Song,~T.~W.; Park,~H.~Y.; Watanabe,~K.; Taniguchi,~T.; others Electrical Interrogation of Thickness-Dependent Multiferroic Phase Transitions in the 2D Antiferromagnetic Semiconductor NiI$_2$. \emph{Advanced Functional Materials} \textbf{2023}, \emph{33}, 2212568\relax
\mciteBstWouldAddEndPuncttrue
\mciteSetBstMidEndSepPunct{\mcitedefaultmidpunct}
{\mcitedefaultendpunct}{\mcitedefaultseppunct}\relax
\EndOfBibitem
\bibitem[Bai \latin{et~al.}(2021)Bai, Zhang, Dun, Zhang, Huang, Zhou, Stone, Kolesnikov, Ye, Batista, \latin{et~al.} others]{Bai2021}
Bai,~X.; Zhang,~S.-S.; Dun,~Z.; Zhang,~H.; Huang,~Q.; Zhou,~H.; Stone,~M.~B.; Kolesnikov,~A.~I.; Ye,~F.; Batista,~C.~D.; others Hybridized quadrupolar excitations in the spin-anisotropic frustrated magnet FeI$_2$. \emph{Nature Physics} \textbf{2021}, \emph{17}, 467--472\relax
\mciteBstWouldAddEndPuncttrue
\mciteSetBstMidEndSepPunct{\mcitedefaultmidpunct}
{\mcitedefaultendpunct}{\mcitedefaultseppunct}\relax
\EndOfBibitem
\bibitem[Kim \latin{et~al.}(2023)Kim, Kim, Park, Kim, Jeong, Ohira-Kawamura, Murai, Nakajima, Chernyshev, Mourigal, \latin{et~al.} others]{Kim2023}
Kim,~C.; Kim,~S.; Park,~P.; Kim,~T.; Jeong,~J.; Ohira-Kawamura,~S.; Murai,~N.; Nakajima,~K.; Chernyshev,~A.; Mourigal,~M.; others Bond-dependent anisotropy and magnon decay in cobalt-based Kitaev triangular antiferromagnet. \emph{Nature Physics} \textbf{2023}, \emph{19}, 1624--1629\relax
\mciteBstWouldAddEndPuncttrue
\mciteSetBstMidEndSepPunct{\mcitedefaultmidpunct}
{\mcitedefaultendpunct}{\mcitedefaultseppunct}\relax
\EndOfBibitem
\bibitem[Kurumaji \latin{et~al.}(2013)Kurumaji, Seki, Ishiwata, Murakawa, Kaneko, and Tokura]{Kurumaji2013}
Kurumaji,~T.; Seki,~S.; Ishiwata,~S.; Murakawa,~H.; Kaneko,~Y.; Tokura,~Y. Magnetoelectric responses induced by domain rearrangement and spin structural change in triangular-lattice helimagnets NiI$_2$ and CoI$_2$. \emph{Physical Review B} \textbf{2013}, \emph{87}, 014429\relax
\mciteBstWouldAddEndPuncttrue
\mciteSetBstMidEndSepPunct{\mcitedefaultmidpunct}
{\mcitedefaultendpunct}{\mcitedefaultseppunct}\relax
\EndOfBibitem
\bibitem[Kuindersma \latin{et~al.}(1981)Kuindersma, Sanchez, and Haas]{Kuindersma1981}
Kuindersma,~S.; Sanchez,~J.; Haas,~C. Magnetic and structural investigations on NiI$_2$ and CoI$_2$. \emph{Physica B+C} \textbf{1981}, \emph{111}, 231--248\relax
\mciteBstWouldAddEndPuncttrue
\mciteSetBstMidEndSepPunct{\mcitedefaultmidpunct}
{\mcitedefaultendpunct}{\mcitedefaultseppunct}\relax
\EndOfBibitem
\bibitem[Song \latin{et~al.}(2022)Song, Occhialini, Erge{\c{c}}en, Ilyas, Amoroso, Barone, Kapeghian, Watanabe, Taniguchi, Botana, \latin{et~al.} others]{Song2022}
Song,~Q.; Occhialini,~C.~A.; Erge{\c{c}}en,~E.; Ilyas,~B.; Amoroso,~D.; Barone,~P.; Kapeghian,~J.; Watanabe,~K.; Taniguchi,~T.; Botana,~A.~S.; others Evidence for a single-layer van der Waals multiferroic. \emph{Nature} \textbf{2022}, \emph{602}, 601--605\relax
\mciteBstWouldAddEndPuncttrue
\mciteSetBstMidEndSepPunct{\mcitedefaultmidpunct}
{\mcitedefaultendpunct}{\mcitedefaultseppunct}\relax
\EndOfBibitem
\bibitem[Billerey \latin{et~al.}(1980)Billerey, Terrier, Mainard, and Pointon]{Billerey1980}
Billerey,~D.; Terrier,~C.; Mainard,~R.; Pointon,~A. Magnetic phase transition in anhydrous NiI$_2$. \emph{Physics Letters A} \textbf{1980}, \emph{77}, 59--60\relax
\mciteBstWouldAddEndPuncttrue
\mciteSetBstMidEndSepPunct{\mcitedefaultmidpunct}
{\mcitedefaultendpunct}{\mcitedefaultseppunct}\relax
\EndOfBibitem
\bibitem[Billerey \latin{et~al.}(1977)Billerey, Terrier, Ciret, and Kleinclauss]{Billerey1977}
Billerey,~D.; Terrier,~C.; Ciret,~N.; Kleinclauss,~J. Neutron diffraction study and specific heat of antiferromagnetic NiI$_2$. \emph{Physics Letters A} \textbf{1977}, \emph{61}, 138--140\relax
\mciteBstWouldAddEndPuncttrue
\mciteSetBstMidEndSepPunct{\mcitedefaultmidpunct}
{\mcitedefaultendpunct}{\mcitedefaultseppunct}\relax
\EndOfBibitem
\bibitem[Ju \latin{et~al.}(2021)Ju, Lee, Kim, Choi, Roh, Son, Park, Kim, Jung, Kim, \latin{et~al.} others]{Ju2021}
Ju,~H.; Lee,~Y.; Kim,~K.-T.; Choi,~I.~H.; Roh,~C.~J.; Son,~S.; Park,~P.; Kim,~J.~H.; Jung,~T.~S.; Kim,~J.~H.; others Possible persistence of multiferroic order down to bilayer limit of van der Waals material NiI$_2$. \emph{Nano letters} \textbf{2021}, \emph{21}, 5126--5132\relax
\mciteBstWouldAddEndPuncttrue
\mciteSetBstMidEndSepPunct{\mcitedefaultmidpunct}
{\mcitedefaultendpunct}{\mcitedefaultseppunct}\relax
\EndOfBibitem
\bibitem[Fumega and Lado(2022)Fumega, and Lado]{Fumega2022}
Fumega,~A.~O.; Lado,~J. Microscopic origin of multiferroic order in monolayer ce{NiI2}. \emph{2D Materials} \textbf{2022}, \emph{9}, 025010\relax
\mciteBstWouldAddEndPuncttrue
\mciteSetBstMidEndSepPunct{\mcitedefaultmidpunct}
{\mcitedefaultendpunct}{\mcitedefaultseppunct}\relax
\EndOfBibitem
\bibitem[Mekata \latin{et~al.}(1992)Mekata, Kuriyama, Ajiro, Mitsuda, and Yoshizawa]{Mekata1992}
Mekata,~M.; Kuriyama,~H.; Ajiro,~Y.; Mitsuda,~S.; Yoshizawa,~H. First-order magnetic transition in CoI$_2$. \emph{Journal of Magnetism and Magnetic Materials} \textbf{1992}, \emph{104}, 859--860\relax
\mciteBstWouldAddEndPuncttrue
\mciteSetBstMidEndSepPunct{\mcitedefaultmidpunct}
{\mcitedefaultendpunct}{\mcitedefaultseppunct}\relax
\EndOfBibitem
\bibitem[Szczerbakow and Durose(2005)Szczerbakow, and Durose]{Szczerbakow2005}
Szczerbakow,~A.; Durose,~K. Self-selecting vapour growth of bulk crystals--Principles and applicability. \emph{Progress in crystal growth and characterization of materials} \textbf{2005}, \emph{51}, 81--108\relax
\mciteBstWouldAddEndPuncttrue
\mciteSetBstMidEndSepPunct{\mcitedefaultmidpunct}
{\mcitedefaultendpunct}{\mcitedefaultseppunct}\relax
\EndOfBibitem
\bibitem[Yan and McGuire(2023)Yan, and McGuire]{Yan2023}
Yan,~J.-Q.; McGuire,~M.~A. Self-selecting vapor growth of transition-metal-halide single crystals. \emph{Physical Review Materials} \textbf{2023}, \emph{7}, 013401\relax
\mciteBstWouldAddEndPuncttrue
\mciteSetBstMidEndSepPunct{\mcitedefaultmidpunct}
{\mcitedefaultendpunct}{\mcitedefaultseppunct}\relax
\EndOfBibitem
\end{mcitethebibliography}

\end{document}